\documentclass[dvips,aoas,preprint]{imsart2}

\usepackage{amsmath,amssymb,amsthm}
\usepackage[final]{graphicx}
\usepackage{float}
\usepackage[round,authoryear]{natbib}
\RequirePackage[colorlinks,citecolor=blue,urlcolor=blue]{hyperref}

\usepackage{verbatim} 
\usepackage{color}
\usepackage{floatpag}
\usepackage{epsfig}

\newcommand{\figdir}{figs} 

\def\uniqueTheta{\theta}

\newcommand{\BF}[1]{\mbox{\boldmath$#1$}}
\newcommand{\DP}[2]{\ensuremath{\mathrm{DP}\!\left({#1}{#2}\right)}}

\newcommand{\symsubsup}[3]{{#1}_{#2}^{(#3)}}
\newcommand{\symsubsupB}[3]{\mathbf{#1}_{#2}^{(#3)}}

\newcommand{\fset}[1]{\ensuremath{\boldsymbol{f}_{\!#1}}}
\floatstyle{boxed}
\newfloat{algorithm}{tbh}{loa}
\floatname{algorithm}{Algorithm}

\begin{document}
	
\begin{frontmatter}

\title{Mixed Membership Models for Time Series}
\runtitle{Mixed Membership Models for Time Series}

\begin{aug}
\author{\fnms{Emily B.} \snm{Fox}\thanksref{wash}\corref{}\ead[label=e1]{ebfox@stat.washington.edu}}
\and
\author{\fnms{Michael I.} \snm{Jordan}\thanksref{Berk}\ead[label=e4]{jordan@stat.berkeley.edu}}

\runauthor{E. Fox and M. Jordan}

\affiliation{University of Washington\thanksmark{wash} and University of California, Berkeley\thanksmark{Berk}}

\address{Emily B. Fox \\
Department of Statistics\\
University of Washington\\
Box 354322\\
Seattle, WA 98195\\
USA\\
\printead{e1}\\
}

\address{Michael I. Jordan\\
Department of Statistics and Department of EECS\\
427 Evans Hall\\
Berkeley, CA 94720\\
USA\\
\printead{e4}\\
}
\end{aug}

\begin{abstract}
	In this article we discuss some of the consequences of the mixed membership perspective on time series analysis. In its most abstract form, a mixed membership model aims to associate an individual \emph{entity} with some set of \emph{attributes} based on a collection of observed data. Although much of the literature on mixed membership models considers the setting in which exchangeable collections of data are associated with each	member of a set of entities, it is equally natural to consider problems in which an entire time series is viewed as an entity and the goal is to characterize the time series in terms of a set of underlying dynamic attributes or \emph{dynamic regimes}. Indeed, this perspective is already present in the classical hidden Markov model, where the dynamic regimes are referred to as ``states,'' and the collection of states realized in a sample path of the underlying process can be viewed as a mixed membership characterization of the observed time series.  Our goal here is to review some of the richer modeling possibilities for time series that are provided by recent developments in the mixed membership framework.
\end{abstract}

\end{frontmatter}

\section{Introduction}

In this article we discuss some of the consequences of the mixed membership perspective
on time series analysis.  In its most abstract form, a mixed membership model aims to associate an individual \emph{entity} with some set of \emph{attributes} based on a collection of observed \emph{data}.  For example, a person (\emph{entity}) can be associated with various defining characteristics (\emph{attributes}) based on observed pairwise interactions with other people (\emph{data}).  Likewise, one can describe a document (\emph{entity}) as comprised of a set of topics (\emph{attributes}) based on the observed words in the document (\emph{data}).  Although much of the literature on mixed membership models considers the setting in which exchangeable collections of data are associated with each member of a set of entities, 
it is equally natural to consider problems in which an entire time series is viewed as an entity and the goal is to characterize the time series in terms of a set of underlying dynamic attributes or \emph{dynamic regimes}.  Indeed, this perspective is already present in the classical hidden Markov model~\citep{Rabiner:89} and switching state-space model~\citep{Kim:94}, where the dynamic regimes are referred to as ``states,'' and the collection of states realized in a sample path of the underlying process can be viewed as a mixed membership characterization of the observed time series.  Our goal here is to review some of the richer modeling possibilities for time series that are provided by recent developments in the mixed membership framework.

Much of our discussion centers around the fact that while in classical time 
series analysis it is commonplace to focus on a single time series, in mixed 
membership modeling it is rare to focus on a single entity (e.g., a single 
document); rather, the goal is to model the way in which multiple entities 
are related according to the overlap in their pattern of mixed membership.  
Thus we take a nontraditional perspective on time series in which the focus 
is on collections of time series.  Each individual time series may be 
characterized as proceeding through a sequence of states, and the focus 
is on relationships in the choice of states among the different time series.

As an example that we review later in this article, consider a multivariate 
time series that arises when position and velocity sensors are placed on the 
limbs and joints of a person who is going through an exercise routine.  
In the specific dataset that we discuss, the time series can be segmented 
into types of exercise (e.g., jumping jacks, touch-the-toes, and twists).  
Each person may select a subset from a library of possible exercise types 
for their individual routine.  The goal is to discover these exercise types 
(i.e., the ``behaviors'' or ``dynamic regimes'') and to identify which person
engages in which behavior, and when.  Discovering and characterizing
``jumping jacks'' in one person's routine should be useful in identifying
that behavior in another person's routine.  In essence, we would like to 
implement a combinatorial form of shrinkage involving subsets of behaviors 
selected from an overall library of behaviors.

Another example arises in genetics, where mixed membership models are referred
to as ``admixture models''~\citep{Pritchard:00}.  Here the goal is to model each individual
genome as a mosaic of marker frequencies associated with different ancestral 
genomes.  If we wish to capture the dependence of nearby markers along the 
genome, then the overall problem is that of capturing relationships among the 
selection of ancestral states along a collection of one-dimensional spatial series. 

One approach to problems of this kind involves a relatively straightforward adaptation of hidden Markov models or other switching state-space models into a Bayesian hierarchical model: transition and emission (or state-space) parameters are chosen from a global prior distribution and each individual time series either uses these global parameters directly or perturbs them further.  This approach in essence involves using a single global library of states, with individual time series differing according to their particular random sequence of states.  This approach is akin to the traditional Dirichlet-multinomial framework that is used in many mixed-membership models.  An alternative is to make use of a beta-Bernoulli framework in which each individual time series is modeled by first selecting a subset of states from a global library and then drawing state sequences from a model defined on that particular subset of states.  We will overview both of these approaches in the remainder of the article.

While much of our discussion is agnostic to the distinction between parametric 
and nonparametric models, our overall focus is on the nonparametric case.  This 
is because the model choice issues that arise in the multiple time series setting
can be daunting, and the nonparametric framework provides at least some initial
control over these issues.  In particular, in a classical state-space setting we 
would need to select the number of states for each individual time series, and
do so in a manner that captures partial overlap in the selected subsets of states
among the time series.  The nonparametric approach deals with these issues as
part of the model specification rather than as a separate model choice procedure.

The remainder of the article is organized as follows.  In Section~\ref{sec:state-space}, we review a set of time series models that form the building blocks for our mixed membership models.  The mixed membership analogy for time series models is aided by relating to a canonical mixed membership model: latent Dirichlet allocation (LDA), reviewed in Section~\ref{sec:LDA}.  Bayesian nonparametric variants of LDA are outlined in Section~\ref{sec:BNP-LDA}.  Building on this background, in Section~\ref{sec:MMtimeseries} we turn our focus to mixed membership in time series.  We first present Bayesian parametric and nonparametric models for single time series in Section~\ref{sec:MSP} and then for collections of time series in Section~\ref{sec:multiple}.  Section~\ref{sec:related} contains a brief survey of related Bayesian and Bayesian nonparametric time series models.
\section{Background}
\label{sec:background}
In this section we provide a brief introduction to some basic terminology from
time series analysis.  We also overview some of the relevant background from mixed
membership modeling, both parametric and nonparametric.

\subsection{State-Space models}
\label{sec:state-space}
The autoregressive (AR) process is a classical model for time series analysis that
we will use as a building block.  An AR model assumes that each observation 
is a function of some fixed number of previous observations plus an uncorrelated 
\emph{innovation}.  Specifically, a linear, time-invariant AR model has the 
following form:
\begin{align}
	y_t = \sum_{i=1}^r a_i y_{t-i} + \epsilon_t,
\end{align}
where $y_t$ represents a sequence of equally spaced observations, $\epsilon_t$ the 
uncorrelated innovations, and $a_i$ the time-invariant autoregressive parameters.  
Often one assumes normally distributed innovations $\epsilon_t \sim \mathcal{N}(0,\sigma^2)$, 
further implying that the innovations are \emph{independent}.    

A more general formulation is that of \emph{linear state space models}, sometimes 
referred to as \emph{dynamic linear models}.  This formulation, which is closely related
to autoregressive moving average processes, assumes that there exists an underlying 
state vector $\BF{x}_t \in \mathbb{R}^n$ such that the past and future of the dynamical 
process $\BF{y}_t \in \mathbb{R}^d$ are conditionally independent.  A linear time-invariant 
state space model is given by
\begin{align}
	\BF{x}_{t} = A\BF{x}_{t-1} + \BF{e}_t \hspace{0.25in} \BF{y}_t = C\BF{x}_{t} + \BF{w}_t, 
\end{align}
where $e_t$ and $w_t$ are independent, zero-mean Gaussian noise processes with covariances 
$\Sigma$ and $R$, respectively.  Here, we assume a \emph{vector-valued} process.  One could 
likewise consider a vector-valued AR process, as we do in Section~\ref{sec:MSP}.

There are several ways to move beyond linear state space models.  One approach is to consider smooth nonlinear functions in place of the matrix multiplication in linear models.  Another approach, which is our focus here, is to consider \emph{regime-switching} models based on a latent sequence of discrete states $\{z_t\}$.  In particular, we consider \emph{Markov switching processes} where the state sequence is modeled as Markovian. If the entire state is a discrete random variable, and the observations $\{y_t\}$ are modeled as being conditionally independent given the discrete state, then we are in the realm of \emph{hidden Markov models} (HMMs)~\citep{Rabiner:89}.  Details of the HMM formulation are expounded upon in Section~\ref{sec:MSP}.

It is also useful to consider hybrid models in which the state contains both 
discrete and continuous components.  We will discuss an important example of this 
formulation---the autoregressive HMM---in Section~\ref{sec:MSP}.  Such models can be 
viewed as a collection of AR models, one for each discrete state.  We will find it 
useful to refer to the discrete states as ``dynamic regimes'' or ``behaviors'' in the setting 
of such models.  Conditional on the value of a discrete state, the model does not 
merely produce independent observations, but exhibits autoregressive behavior.

\subsection{Latent Dirichlet Allocation}
\label{sec:LDA}
In this section, we briefly overview the latent Dirichlet allocation (LDA) model~\citep{Blei:03} 
as a a canonical example of a mixed membership model.  We use the language of ``documents,''
``topics,'' and ``words.''  In contrast to hard-assignment predecessors that assumed each 
document was associated with a single topic category, LDA aims to model each document as a 
mixture of topics.  Throughout this article, when describing a mixed membership model, we 
seek to define some observed quantity as an \emph{entity} that is allowed to be associated
with, or have \emph{membership} characterized by, multiple \emph{attributes}.  For LDA, 
the entity is a \emph{document} and the attributes are a set of possible \emph{topics}.  
Typically, in a mixed membership model each entity represents a set of observations and a 
key question is what structure is imposed on these observations.  For LDA, each document is 
a collection of observed \emph{words} and the model makes a simplifying \emph{exchangeability} 
assumption in which the ordering of words is ignored.  

Specifically, LDA associates each document~$d$ with a latent distribution over the possible topics, $\pi^{(d)}$, and each topic $k$ is associated with a distribution over words in the vocabulary, $\theta_k$.  Each word~$w_i^{(d)}$ is then generated by first selecting a topic from the document-specific topic distribution, and then selecting a word from the topic-specific word distribution.  

Formally, the standard LDA model with $K$ topics, $D$ documents, and $N_d$ words per document~$d$ is given as
\begin{equation}
\begin{aligned}
	\theta_k &\sim \mbox{Dir}(\eta_1,\dots,\eta_V) \qquad k=1,\dots, K\\
	\pi^{(d)}  &\sim \mbox{Dir}(\beta_1,\dots,\beta_K) \qquad d = 1,\dots D\\
	z_i^{(d)} \mid \pi^{(d)} &\sim \pi^{(d)} \qquad d = 1,\dots D, \,\, i = 1,\dots,N_d\\
	w_i^{(d)} \mid \{\theta_k\}, z_i^{(d)} &\sim \theta_{z_i^{(d)}} \qquad d = 1,\dots D, \,\, i = 1,\dots,N_d.
\end{aligned}
\label{eqn:LDA}
\end{equation}
Here $z_i^{(d)}$ is a topic indicator variable associated with observed word $w_i^{(d)}$, indicating which topic~$k$ generated this $i$th word in document~$d$.  In expectation, for each document $d$ we have $E[\pi_k^{(d)}\mid \beta] = \beta_k$.  That is, the expected topic proportions for each document are identical a priori.
\subsection{Bayesian Nonparametric Mixed Membership Models}
\label{sec:BNP-LDA}
The LDA model of Equation~\eqref{eqn:LDA} assumes a finite number of topics $K$.  Bayesian nonparametric methods allow for extensions to models with an unbounded number of topics.  That is, in the mixed membership analogy, each entity can be associated with a potentially countably infinite number of attributes.  We review two such approaches: one based on the hierarchical Dirichlet process~\citep{Teh:06} and the other based on the beta process~\citep{Hjort:90,Thibaux:07}.  In the latter case, the association of entities with attributes is directly modeled as \emph{sparse}.

\paragraph{Hierarchical Dirichlet Process Topic Models}
To allow for a countably infinite collection of topics, in place of finite-dimensional topic-distributions $\pi^{(d)} = [\pi_1^{(d)}, \, \dots, \, \pi_K^{(d)}]$ as specified in Equation~\eqref{eqn:LDA}, one wants to define 
distributions whose support lies on a countable set, $\pi^{(d)} = [\pi_1^{(d)}, \,\pi_2^{(d)}, \, \dots]$. 

\begin{figure}
	[t!] \centering 
\includegraphics[width=.5\textwidth]{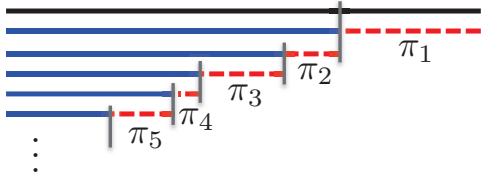}
\caption{Pictorial representation of the stick-breaking construction of the Dirichlet process.} 
\label{fig:stick}
\end{figure}

The \emph{Dirichlet process} (DP), denoted by $\DP{\alpha}{H}$, provides a distribution over countably infinite discrete probability measures
\begin{align}
	G = \sum_{k=1}^\infty \pi_{k}\delta_{\theta_k} \hspace{0.25in} \theta_k \sim H
	\label{eqn:stickDP}
\end{align}
defined on a parameter space $\Theta$ with base measure $H$. The mixture weights are sampled via a \emph{stick-breaking construction} \citep{Sethuraman:94}:
\begin{align}
\pi_k = \nu_k\prod_{\ell=1}^{k-1}(1-\nu_\ell) \qquad
\nu_k \sim \mbox{Beta}(1,\alpha).
\end{align}
This can be viewed as dividing a unit-length stick into lengths given by the weights $\pi_k$: the $k^{th}$ weight is a random proportion $v_k$ of the remaining stick after the first $(k-1)$ weights have been chosen. We denote this distribution by $\pi \sim \mbox{GEM}(\alpha)$. See Figure~\ref{fig:stick} for a pictorial representation of this process.

Drawing indicators $z_i \sim \pi$, one can integrate the underlying random stick-breaking measure $\pi$ to examine the predictive distribution of $z_i$ conditioned on a set of indicators $z_1,\dots,z_{i-1}$ and the DP concentration parameter $\alpha$.  The resulting sequence of partitions is described via the \emph{Chinese restaurant process} (CRP)~\citep{Pitman:02}, which provides insight into the clustering properties induced by the DP.

For the LDA model, recall that each $\theta_k$ is a draw from a Dirichlet distribution (here denoted generically by $H$) and defines a distribution over the vocabulary for topic~$k$.  To define a model for multiple 
documents, one might consider independently sampling $G^{(d)} \sim \DP{\alpha}{H}$ for each document $d$, where each of these random measures is of the form $G^{(d)} = \sum_{k=1}^\infty \pi_k^{(d)}\delta_{\theta_k^{(d)}}$.  Unfortunately, the topic-specific word distribution for document $d$, $\theta_k^{(d)}$, is necessarily different from that of document $d'$, $\theta_k^{(d')}$, since each are independent draws from the base measure $H$.  This is clearly not a desirable model---in a mixed membership model we want the parameter that describes each attribute (\emph{topic}) to be shared between entities (\emph{documents}). 
\begin{figure}
	[t!] \centering 
	\begin{tabular}{cc}
		\includegraphics[height=2in]{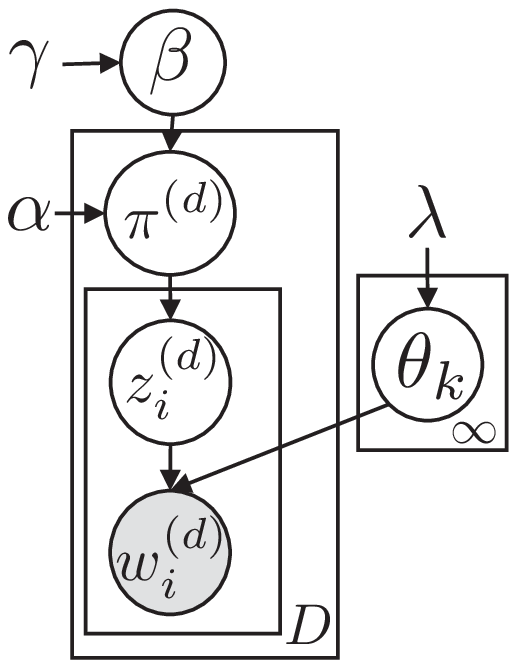} & \includegraphics[height=2in]{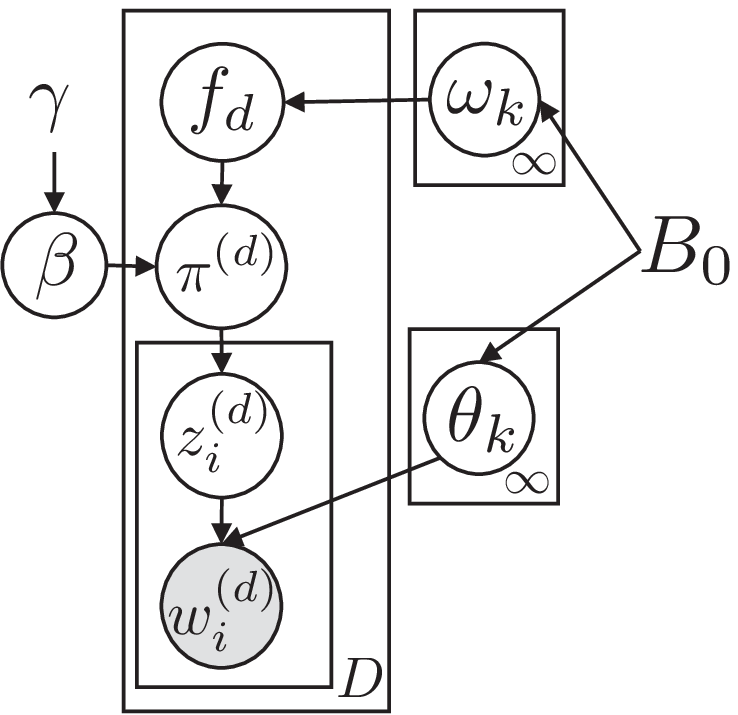}\\
		(a) & (b)
	\end{tabular}
	\caption{Graphical model of the (a) HDP-based and (b) beta-process-based topic model. The HDP-LDA model specifies a global topic distribution $\beta \sim \mbox{GEM}(\gamma)$ and draws document-specific topic distributions as $\pi^{(d)} \mid \beta \sim \DP{\alpha}{\beta}$.  Each word $w_i^{(d)}$ in document $d$ is generated by first drawing a topic-indicator $z_i^{(d)} \mid \pi^{(d)} \sim \pi^{(d)}$ and then drawing from the topic-specific word distribution: $w_i^{(d)} \mid \{\theta_k\},z_i^{(d)} \sim \theta_{z_i^{(d)}}$.  The standard LDA model arises as a special case when $\beta$ is fixed to a finite measure $\beta = [\beta_1,\dots,\beta_K]$.  The beta process model specifies a collection of \emph{sparse} topic distributions.  Here, the beta process measure $\mbox{$B \sim \mbox{BP}(1,B_0)$}$ is represented by its masses $\omega_k$ and locations $\theta_k$, as in Equation~\eqref{eqn:CRM}. The features are then conditionally independent draws $\mbox{$f_{dk} \mid \omega_k \sim \mbox{Bernoulli}(\omega_k)$}$, and are used to define document-specific topic distributions $\mbox{$\symsubsup{\pi}{j}{d} \mid \fset{d},\beta \sim \mbox{Dir}(\beta \otimes \fset{d})$}$. Given the topic distributions, the generative process for the topic-indicators $z_i^{(d)}$ and words $w_i^{(d)}$ is just as in the HDP-LDA model.} \label{fig:LDA} 
\end{figure}

\begin{figure}
	[t!] \centering 
	\includegraphics[width=0.9\columnwidth]{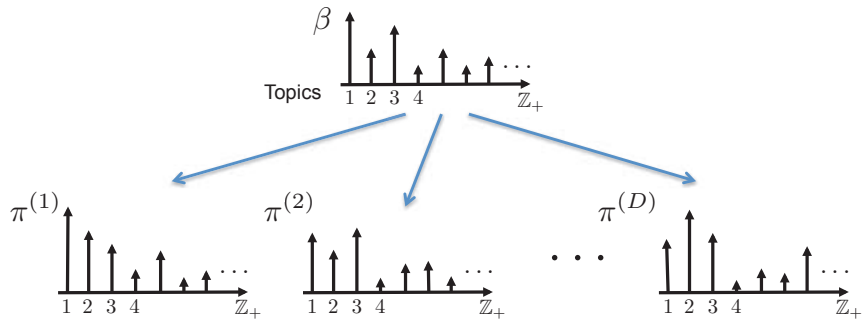}
	\caption{Illustration of the coupling of the document-specific topic distributions $\pi^{(d)}$ via the global stick-breaking distribution $\beta$.  Each topic distribution has countably infinite support and, in expectation, $E[\pi^{(d)}\mid \beta] = \beta_k$.} \label{fig:LDAstick} 
\end{figure}

One method of sharing parameters $\theta_k$ between documents while allowing for document-specific topic weights $\pi^{(d)}$ is to employ the \emph{hierarchical Dirichlet process} (HDP)~\citep{Teh:06}.  The HDP defines a shared set of parameters by drawing $\theta_k$ independently from $H$.  The weights are then specified as 
\begin{align}
\beta \sim \mbox{GEM}(\gamma) \qquad \pi^{(d)}\mid \beta \sim \DP{\alpha}{\beta}.
\label{eqn:stickHDP}
\end{align}
Coupling this prior to the likelihood used in the LDA model, we obtain a model that we refer to as \emph{HDP-LDA}.  See Figure~\ref{fig:LDA}(a) for a graphical model representation, and Figure~\ref{fig:LDAstick} for an illustration of the coupling of document-specific topic distributions via the global stick-breaking distribution $\beta$.  Letting $G^{(d)} = \sum_{k=1}^\infty \pi^{(d)}_k \delta_{\theta_k}$ and $G^{(0)} = \sum_{k=1}^\infty \beta_k \delta_{\theta_k}$, one can show that the specification of Equation~\eqref{eqn:stickHDP} is equivalent to defining a hierarchy of Dirichlet processes~\citep{Teh:06}:
\begin{align}
G^{(0)} \sim \DP{\gamma}{H} \qquad G^{(d)} \mid G^{(0)} \sim \DP{\alpha}{G^{(0)}}.
\label{eqn:HDP}
\end{align}
Thus the name \emph{hierarchical} Dirichlet process.  Note that there are many possible alternative formulations one could have considered to generate different countably infinite weights $\pi^{(d)}$ with shared atoms $\theta_k$.  
The HDP is a particularly simple instantiation of such a model that has appealing theoretical and computational properties due to its interpretation as a hierarchy of Dirichlet processes.

Via the construction of Equation~\eqref{eqn:stickHDP}, we have that $E[\pi_k^{(d)} \mid \beta] = \beta_k$.  That is, all of the document-specific topic distributions are centered around the same stick-breaking weights $\beta$.
		
\paragraph{Beta-Bernoulli Process Topic Models}
The HDP-LDA model defines countably infinite topic distributions $\pi^{(d)}$ in which every topic~$k$ has positive mass $\pi_k^{(d)} > 0$ (see Figure~\ref{fig:LDAstick}).  This implies that each entity (\emph{document}) is associated with infinitely many attributes (\emph{topics}).  In practice, however, for any finite length document~$d$ only a finite subset of the topics will be present.  The HDP-LDA model implicitly provides such \emph{attribute counts} through the assignment of words $w_i^{(d)}$ to topics via the indicator variables $z_i^{(d)}$.

As an alternative representation that more directly captures the inherent sparsity of association between documents and topics, one can consider \emph{feature-based} Bayesian nonparametric variants of LDA via the \emph{beta-Bernoulli process}, such as in the \emph{focused topic model} of~\cite{Williamson:10}.  (A precursor to this model
was presented in the time series context by~\cite{Fox:NIPS09}, and is discussed in Section~\ref{sec:multiple}.)  
In such models, each document is endowed with an infinite-dimensional binary feature vector that indicates which topics are associated with the given document.  In contrast to HDP-LDA, this formulation directly allows each document to be represented as a \emph{sparse} mixture of topics.  That is, there are only a few topics that have positive probability of appearing in any document.  

Informally, one can think of the \emph{beta process} (BP)~\citep{Hjort:90,Thibaux:07} as defining an 
infinite set of coin-flipping probabilities and a Bernoulli process realization as corresponding 
to the outcome from an infinite coin-flipping sequence based on the beta-process-determined 
coin-tossing probabilities.  The set of resulting \emph{heads} indicate the set of selected 
\emph{features}, and implicitly defines an infinite-dimensional feature vector.  The properties 
of the beta process induce sparsity in the feature space by encouraging sharing of features among the Bernoulli process realizations.   

More formally, let $\fset{d} = [f_{d1}, f_{d2}, \ldots]$ be an infinite-dimensional feature vector associated with document $d$, where $f_{dk}=1$ if and only if document~$d$ is associated with topic~$k$.  The beta process, denoted $\mbox{BP}(c,B_0)$, provides a distribution on measures
\begin{align}
	B = \sum_{k=1}^\infty \omega_k\delta_{\theta_k},
	\label{eqn:CRM}
\end{align}
with $\omega_k \in (0,1)$.  We interpret $\omega_k$ as the feature-inclusion probability for feature $k$ (e.g., the $k$th topic in an LDA model).  This $k$th feature is associated with parameter $\theta_k$.

\begin{figure}[t]
  \centering
\begin{tabular}{cc}
  \includegraphics[width=0.5\columnwidth]{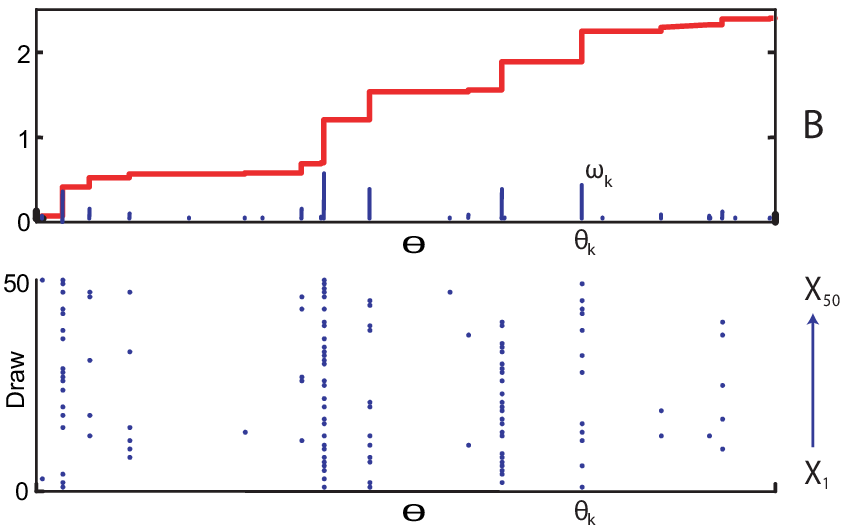}
& \includegraphics[width =
0.4\columnwidth]{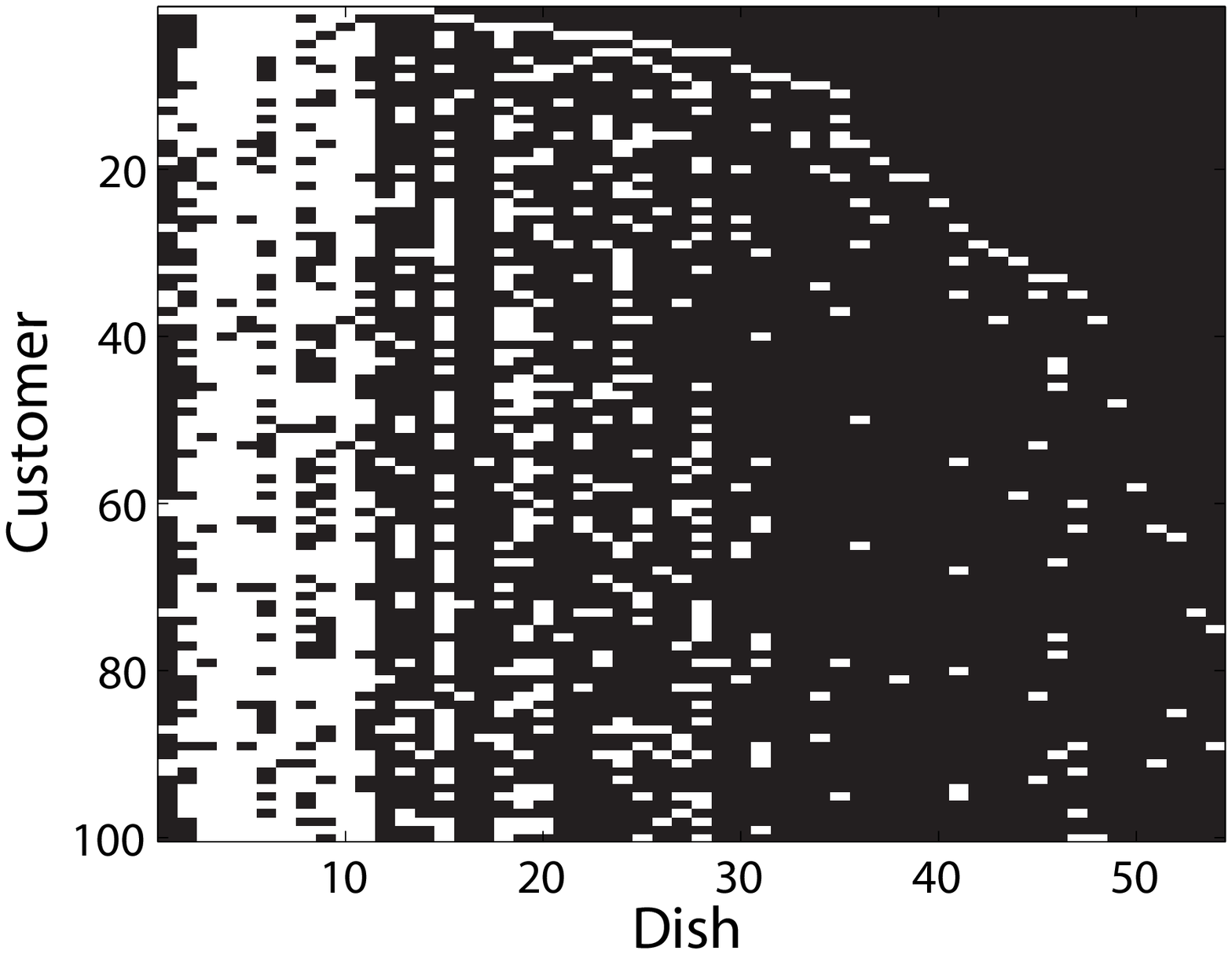}\\
(a) & (b)
\end{tabular}
  \caption[A draw from a beta process, and associated Bernoulli realizations, along with a realization from the Indian buffet process.]{(a) \textit{Top}: A draw $B$ from a beta process is shown in blue, with the corresponding cumulative distribution in red. \textit{Bottom}: 50 draws $X_i$ from a Bernoulli process using the beta process realization. Each blue dot corresponds to a coin-flip at that atom in $B$ that came up heads. (b) An image of a feature matrix associated with a realization from an Indian buffet process with $\alpha=10$. Each row corresponding to a different customer, and each column a different dish.  White indicates a chosen feature.} \label{fig:BPBePrealizations}
\end{figure}

The collection of points $\{\theta_k,\omega_k\}$ are a draw from a non-homogeneous Poisson process with rate $\nu(d\omega, d\theta) = c\omega^{-1}(1 - \omega)^{c-1}d\omega B_0(d\theta)$ defined on the product space $\Theta \otimes [0,1]$.  Here, $c > 0$ and $B_0$ is a base measure with total mass $B_0(\Theta) = \alpha$.  Since the rate measure $\eta$ has infinite mass, the draw from the Poisson process yields an infinite collection of points, as in Equation~\eqref{eqn:CRM}.  For an example realization and its associated cumulative distribution, see Figure~\ref{fig:BPBePrealizations}.  One can also interpret the beta process as the limit of a finite model with $K$ features:
\begin{align}
	B_K = \sum_{k=1}^K \omega_k\delta_{\theta_k} \qquad
	\omega_k \sim \mbox{Beta}\left(\frac{c\alpha}{K},c(1-\frac{\alpha}{K})\right) \qquad
	\theta_k \sim \alpha^{-1}B_0.
\end{align}
In the limit as $K\rightarrow \infty$, $B_K \rightarrow B$ and one can define stick-breaking constructions 
analogous to those in the Dirichlet process~\citep{Paisley:10,Paisley:11}.  

For each feature $k$, we independently sample
\begin{align}
f_{dk} \mid \omega_k \sim \mbox{Bernoulli}(\omega_k).
\end{align}
That is, with probability $\omega_k$, topic~$k$ is associated with document $d$.  One can visualize this process as walking along the atoms of the discrete beta process measure $B$ and, at each atom $\theta_k$, flipping a coin with probability of heads given by $\omega_k$.  More formally, setting $X_d = \sum_{k=1}^\infty f_{dk}\delta_{\theta_k}$, this process is equivalent to sampling $X_d$ from a \emph{Bernoulli process} with base measure $B$: $X_d\mid B \sim \mbox{BeP}(B)$.  Example realizations are shown in Figure~\ref{fig:BPBePrealizations}(a).

The characteristics of this beta-Bernoulli process define desirable traits for a Bayesian nonparametric featural model: we have a countably infinite collection of coin-tossing probabilities (one for each of our infinite number of features) defined by the beta process, but only a sparse, finite subset are active in any Bernoulli process realization.  In particular, one can show that $B$ has finite expected mass implying that there are only a finite number of successes in the infinite coin-flipping sequence that defines $X_d$.  Likewise, the sparse set of features active in $X_d$ are likely to be similar to those of $X_{d'}$ (an independent draw from $\mbox{BeP}(B)$), though variability is clearly possible.  Finally, the beta process is conjugate to the Bernoulli process~\citep{Kim:99}, which implies that one can analytically marginalize the latent random beta process measure $B$ and examine the predictive distribution of $\fset{d}$ given $\fset{1},\dots,\fset{d-1}$ and the concentration parameter $\alpha$.  As established 
by~\cite{Thibaux:07}, the marginal distribution on the $\{\fset{d}\}$ obtained from the beta-Bernoulli 
process is the \emph{Indian buffet process} (IBP) of~\cite{GriffithsGhahramani:05}, just as the marginalization
of the Dirichlet-multinomial process yields the Chinese restaurant process.  The IBP can be useful in
developing posterior inference algorithms and a significant portion of the literature is written in 
terms of the IBP representation.

\begin{figure}
	[t!] \centering 
	\includegraphics[width=0.9\columnwidth]{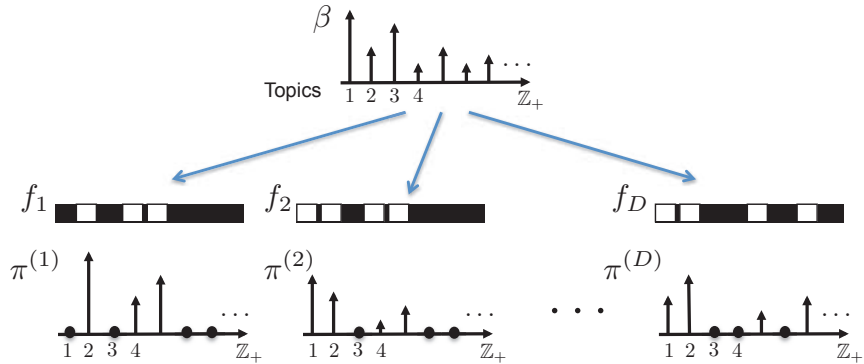}
	\caption{Illustration of generating the \emph{sparse} document-specific topic distributions $\pi^{(d)}$ via the beta process specification.  Each document's binary feature vector $\fset{d}$ limits the support of the topic distribution to the sparse set of selected topics.  The non-zero components are Dirichlet distributed with hyperparmeters given by the corresponding subset of $\beta$.  See Equation~\eqref{eqn:BP-LDA}.} \label{fig:LDAsparsestick} 
\end{figure}
Returning to the LDA model, one can obtain the focused topic model of~\cite{Williamson:10} within
the beta-Bernoulli process framework as follows:
\begin{equation}
\begin{aligned}
	B &\sim \mbox{BP}(1,B_0)\\
	X_d \mid B &\sim \mbox{BeP}(B) \quad d = 1,\dots D\\
	\pi^{(d)} \mid f_d,\beta &\sim \mbox{Dir}(\beta\otimes \fset{d}) \quad d = 1,\dots D,
\end{aligned}
\label{eqn:BP-LDA}
\end{equation}
where~\cite{Williamson:10} treat $\beta$ as random according to $\beta_k \sim \mbox{Gamma}(\gamma,1)$. Here, $\fset{d}$ is the feature vector associated with $X_d$ and $\mbox{Dir}(\beta\otimes \fset{d})$ represents a Dirichlet distribution defined solely over the components indicated by $\fset{d}$, with hyperparameters the corresponding subset of $\beta$.  This implies that $\pi^{(d)}$ is a distribution with positive mass only on the \emph{sparse} set of selected topics.  See Figure~\ref{fig:LDAsparsestick}.  Given $\pi^{(d)}$, the $z_i^{(d)}$ and $w_i^{(d)}$ are generated just as in Equation~\eqref{eqn:LDA}.  As before, we take $\theta_k \sim \mbox{Dir}(\eta_1,\dots,\eta_V)$.  The graphical model is depicted in Figure~\ref{fig:LDA}(b).
%
%
%
\section{Mixed Membership in Time Series}
\label{sec:MMtimeseries}
Building on the background provided in Section~\ref{sec:background}, we can now explore how ideas of mixed membership models can be used in the time series setting.  Our particular focus is on time series that can be well described using \emph{regime-switching} models.  For example, stock returns might be modeled as switches between regimes of volatility or an EEG recording between spiking patterns dependent on seizure type.  For the exercise routines scenario, people switch between a set of actions such as jumping jacks, side twists, and so on.  In this section, we present a set of regime-switching models for describing such datasets, and show how one can interpret the models as providing a form of mixed membership for time series.

To form the mixed membership interpretation, we build off of the canonical example of LDA from Section~\ref{sec:LDA}.  Recall that for LDA, the entity of interest is a \emph{document} and the set of attributes are the possible \emph{topics}.  Each document is then modeled as having membership in multiple topics (i.e., \emph{mixed membership}).  For time series analysis, the equivalent analogy is that the entity is the \emph{time series} $\{\BF{y}_t: t=1,\dots,T\}$, which we denote compactly by $\BF{y}_{1:T}$.  Just as a document is a collection of observed \emph{words}, a time series is a sequence of observed \emph{data points} of various forms depending upon the application domain.  We take the attributes of a time series to be the collection of \emph{dynamic regimes} (e.g., jumping jacks, arm 
circles, etc.).  Our mixed membership time series model associates a single time series with a collection 
of dynamic regimes.  However, unlike in text analysis, it is unreasonable to assume a \emph{bag of words} representation for time series since the ordering of the data points is fundamental to the description of each dynamic regime.  

The central defining characteristics of a mixed membership time series model are (i) the model used to describe each dynamic regime, and (ii) the model used to describe the switches between regimes.  In Section~\ref{sec:MSP} and 
in Section~\ref{sec:HDP-HMM} we choose one switching model and explore multiple choices for the dynamic regime model.  Another interesting question explored in Section~\ref{sec:multiple} is how to jointly model multiple time series.  This question is in direct analogy to the ideas behind the analysis of a \emph{corpus} of documents in LDA.
%
%
\subsection{Markov Switching Processes as a Mixed Membership Model}
\label{sec:MSP}
A flexible yet simple regime-switching model for describing a single time series with such patterned behaviors is the class of \emph{Markov switching processes}.  These processes assume that the time series can be described via Markov transitions between a set of latent dynamic regimes which are individually modeled via temporally independent or linear dynamical systems.  Examples include the hidden Markov model (HMM), switching vector autoregressive (VAR) process, and switching linear dynamical system (SLDS)\footnote{These processes are sometimes referred to as \emph{Markov jump-linear systems} (MJLS) within the control theory community.}.  These models have proven useful in such diverse fields as speech recognition, econometrics, neuroscience, remote target tracking, and human motion capture.
\subsubsection*{Hidden Markov Models}
The hidden Markov model, or \emph{HMM}, is a class of doubly stochastic processes based on an underlying, discrete-valued state sequence that is modeled as Markovian~\citep{Rabiner:89}. Conditioned on this state sequence, the model assumes that the observations, which may be discrete or continuous valued, are independent. Specifically, let $z_t$ denote the state, or \emph{dynamic regime}, of the Markov chain at time~$t$ and let $\pi_j$ denote the state-specific \emph{transition distribution} for state $j$. Then, the Markovian structure on the state sequence dictates that
\begin{align}
z_t\mid z_{t-1} \sim \pi_{z_{t-1}}. \label{eqn:HMMstate}
\end{align}
Given the state $z_t$, the observation $y_t$ is a conditionally independent emission
\begin{align}
y_t \mid \{\theta_j\}, z_t \sim F(\theta_{z_t})
\label{eqn:HMMobs}
\end{align}
for an indexed family of distributions $F(\cdot)$.  Here, $\theta_j$ are the \emph{emission parameters} for state~$j$. 

A Bayesian specification of the HMM might further assume
\begin{align}
	\pi_j \sim \mbox{Dir}(\beta_1,\dots,\beta_K) \qquad \theta_j \sim H	
\end{align}
independently for each HMM state $j=1,\dots, K$.

The HMM represents a simple example of a mixed membership model for time series: a given time series (\emph{entity}) is modeled as having been generated from a collection of dynamic regimes (\emph{attributes}), each with different mixture weights.  The key component of the HMM, which differs from standard mixture models such as in LDA, is the fact that there is a Markovian structure to the assignment of data points to mixture components (i.e., dynamic regimes).  In particular, the probability that observation $y_t$ is generated from the dynamic regime associated with state $j$ (via an assignment $z_t=j$) is dependent upon the previous state $z_{t-1}$.  As such, the mixing proportions for the time series are defined by the transition matrix $P$ with rows $\pi_j$.  This is in contrast to the LDA model in which the mixing proportions for a given document are simply captured by a single vector of weights.
\subsubsection*{Switching VAR Processes}
The modeling assumption of the HMM that observations are conditionally independent given the latent state sequence is often insufficient in capturing the temporal dependencies present in many datasets.  Instead, one can assume that the observations have conditionally \emph{linear} dynamics.  The latent HMM state then models switches between a set of such linear models in order to capture more complex dynamical phenomena.  We restrict our attention in this article to switching vector autoregressive (VAR) processes, or \emph{autoregressive HMMs} (AR-HMMs), which are broadly applicable in many domains while maintaining a number of simplifying properties that make them a practical choice computationally.  

We define an AR-HMM, with switches between order-$r$ vector autoregressive processes~\footnote{We denote an order-$r$ VAR process by VAR($r$).}, as
\begin{equation}
\begin{aligned}
\BF{y}_t &= \sum_{i=1}^r A_{i,z_t}\BF{y}_{t-i} + \BF{e}_t(z_t),
\end{aligned}
\label{eqn:SVAR}
\end{equation}
where $z_t$ represents the HMM latent state at time $t$, and is defined as in Equation~\eqref{eqn:HMMstate}.  The state-specific additive noise term is distributed as $\BF{e}_t(z_t) \sim \mathcal{N}(0,\Sigma_{z_t})$.  We refer to $\BF{A}_k = \{A_{1,k},\dots,A_{r,k}\}$ as the set of \emph{lag matrices}.  Note that the standard HMM with Gaussian emissions arises as a special case of this model when $\BF{A}_{k}=\BF{0}$ for all~$k$.
\subsection{Hierarchical Dirichlet Process HMMs}
\label{sec:HDP-HMM}
In the HMM formulation described so far, we have assumed that there are $K$ possible different dynamical regimes.  This begs the question: what if this is not known, and what if we would like to allow for new dynamic regimes to be added as more data are observed?  In such scenarios, an attractive approach is to appeal to Bayesian nonparametrics.  Just as the hierarchical Dirchlet process (HDP) of Section~\ref{sec:BNP-LDA} allowed for a collection of countably infinite topic distributions to be defined over the same set of topic parameters, one can employ the HDP to define an HMM with a set of countably infinite transition distributions defined over the same set of HMM emission parameters.  

In particular, the HDP-HMM of~\cite{Teh:06} defines 
\begin{align}
	\beta \sim \mbox{GEM}(\gamma) \qquad \pi_j\mid \beta \sim \DP{\alpha}{\beta} \qquad \theta_j \sim H.
\end{align}
The evolution of the latent state $z_t$ and observations $y_t$ are just as in Equations~\eqref{eqn:HMMstate} and \eqref{eqn:HMMobs}.  Informally, the Dirichlet process part of the HDP allows for this unbounded state space and encourages the use of only a spare subset of these HMM states.  The hierarchical layering of Dirichlet processes ties together the state-specific transition distribution (via $\beta$), and through this process, creates a \emph{shared} sparse state space.

The induced predictive distribution for the HDP-HMM state $z_t$, marginalizing the transition distributions $\pi_j$, is known as the \emph{infinite HMM} urn model~\citep{Beal:02}.  In particular, the HDP-HMM of~\cite{Teh:06} provides an interpretation of this urn model in terms of an underlying collection of linked random probability measures.  However, the HDP-HMM omits the self-transition bias of the infinite HMM and instead assumes that each transition distribution $\pi_j$ is \emph{identical} in expectation ($E[\pi_{jk}\mid \beta] = \beta_k$), implying that there is no differentiation between self-transitions and moves between different states.  When modeling data with state persistence, as is common in most real-world datasets, the flexible nature of the HDP-HMM prior places significant mass on state sequences with unrealistically fast dynamics.

\begin{figure}
	[t!] \centering 
	\begin{tabular}{cc}
		\includegraphics[width=0.48\columnwidth]{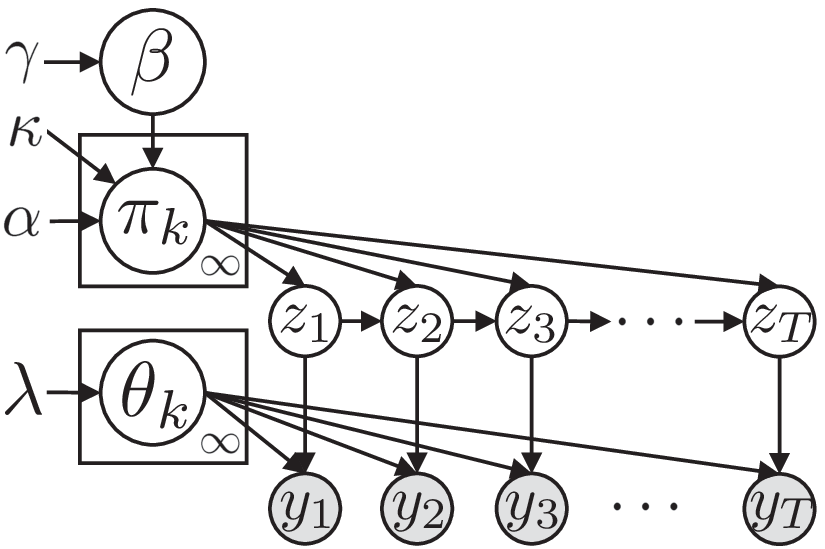} & \includegraphics[width=0.48\columnwidth]{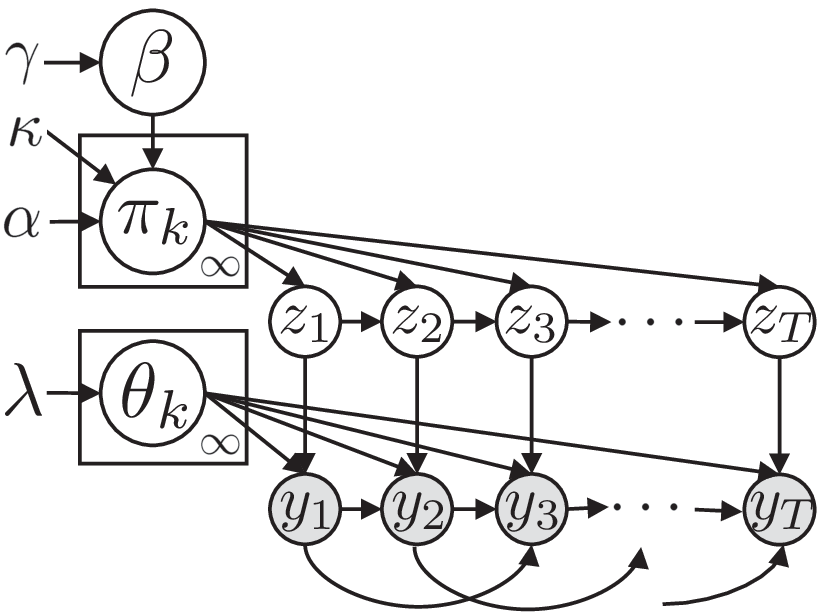}\\
		(a) & (b)
	\end{tabular} 
	\caption{Graphical model of (a) the sticky HDP-HMM and (b) an HDP-based AR-HMM.  In both cases, the state evolves as $z_{t+1}\mid \{\pi_k\},z_t \sim \pi_{z_t}$, where $\pi_k\mid \beta \sim \mbox{DP}(\alpha\beta+\kappa\delta_k)$ and $\beta \sim \mbox{GEM}(\gamma)$.  For the sticky HDP-HMM, the observations are generated as $y_t\mid \{\uniqueTheta_k\},z_t \sim F(\uniqueTheta_{z_t})$ whereas the HDP-AR-HMM assumes conditionally VAR dynamics as in Equation~\eqref{eqn:SVAR}, specifically in this case with order $r=2$.} \label{fig:stickyHDPHMM} 
\end{figure}

To better capture state persistence, the \emph{sticky} HDP-HMM of~\cite{Fox:ICML08,Fox:AOAS11} restores the self-transition parameter of the infinite HMM of~\cite{Beal:02} and specifies
\begin{align}
	\beta \sim \mbox{GEM}(\gamma) \qquad
	\pi_j\mid \beta \sim \mbox{DP}(\alpha\beta + \kappa\delta_j) \qquad \theta_j \sim H,
\end{align}
where $(\alpha\beta + \kappa\delta_j)$ indicates that an amount $\kappa > 0 $ is added to the $j$th component of $\alpha\beta$.  In expectation,
\begin{align}
E[\pi_{jk} \mid \beta, \kappa] = \frac{\alpha\beta_k + \kappa\delta(j,k)}{\alpha+\kappa}.
\label{eqn:stickyMean}
\end{align}
Here, $\delta(j,k)$ is the discrete Kronecker delta.  From Equation~\eqref{eqn:stickyMean}, we see that the expected transition distribution has weights which are a convex combination of the global weights defined by $\beta$ and state-specific weight defined by the sticky parameter $\kappa$.  When $\kappa=0$, the original HDP-HMM of~\cite{Teh:06} is recovered.  The graphical model for the sticky HDP-HMM is displayed in Figure~\ref{fig:stickyHDPHMM}(a).

One can also consider sticky HDP-HMMs with Dirichlet process mixture of Gaussian emissions~\citep{Fox:AOAS11}.  Recently, HMMs with Dirichlet process emissions were also considered in~\cite{Yau:11}, along with efficient sampling algorithms for computations.  Building on the sticky HDP-HMM framework, one can similarly consider HDP-based variants of the switching VAR process and switching linear dynamical system, such as represented in Figure~\ref{fig:stickyHDPHMM}(b); see~\cite{Fox:IEEE11} for further details.  For the \emph{HDP-AR-HMM},~\cite{Fox:IEEE11} consider methods that allow for switching between VAR processes of unknown and potentially variable order.
\subsection{A Collection of Time Series} 
\label{sec:multiple}
In the mixed membership time series models considered thus far, we have assumed that we are interested in the dynamics of a single (potentially multivariate) time series.  However, as in LDA where one assumes a \emph{corpus} of documents, in a growing number of fields the focus is on making inferences based on a \emph{collection} of related time series.  One might monitor multiple financial indices, or collect EEG data from a given patient at multiple non-contiguous epochs.  Recalling the exercise routines example, one might have a dataset consisting of multiple time series obtained from multiple individuals, each of whom performs some subset of exercise types.  In this scenario, we would like to take advantage of the overlap between individuals, such that if a ``jumping jack'' behavior is discovered in the time series for one individual then it can be used in modeling the data for other individuals.  More generally, one would like to discover and model the dynamic regimes that are shared among several related time series.  The benefits of such joint modeling are twofold: we may more robustly estimate representative dynamic models in the presence of limited data, and we may also uncover interesting relationships among the time series.
	
Recall the basic finite HMM of Section~\ref{sec:MSP} in which the transition matrix $P$ defined the dynamic regime mixing proportions for a given time series.  To develop a mixed membership model for a \emph{collection} of time series, we again build on the LDA example.  For LDA, the document-specific mixing proportions over topics are specified by $\pi^{(d)}$.  Analogously, for each time series $y_{1:T_d}^{(d)}$, we denote the time-series specific transition matrix as $P^{(d)}$ with rows $\pi_j^{(d)}$.  That is, for time series~$d$, $\pi_j^{(d)}$ denotes the transition distribution from state~$j$ to each of the $K$ possible next states.  Just as LDA couples the document-specific topic distributions $\pi^{(d)}$ under a common Dirichlet prior, we can couple the \emph{rows} of the transition matrix as
\begin{align}
	\pi_j^{(d)} \sim \mbox{Dir}(\beta_1,\dots,\beta_K).
\end{align}
A similar idea holds for extending the HDP-HMM to collections of time series.  In particular, we can specify
\begin{align}
	\beta \sim \mbox{GEM}(\gamma) \qquad
	\pi_j^{(d)}\mid \beta \sim \DP{\alpha}{\beta}.
\label{eqn:HDPCorporaHierarchy}
\end{align}
Analogously to LDA, both the finite and infinite HMM specifications above imply that the expected transition distributions are identical between time series ($E[\pi_j^{(d)}\mid \beta] = E[\pi_j^{(d')}\mid \beta]$).  Here, however, the expected transition distributions are also identical between rows of the transition matrix.  

To allow for state-specific variability in the expected transition distribution, one could similarly couple sticky HDP-HMMs, or consider a finite variant of the model via the weak-limit approximation (see~\cite{Fox:AOAS11} for details on finite truncations).  Alternatively, one could independently center each row of the time-series-specific transition matrix around a state-specific distribution.  For the finite model,
\begin{align}
	\pi_j^{(d)}\mid \beta_j \sim \mbox{Dir}(\beta_{j1},\dots,\beta_{jK}).
\end{align}
For the infinite model, such a specification is more straightforwardly presented in terms of the Dirichlet random measures.  Let $G^{(d)}_j = \sum \pi_j^{(d)}\delta_{\theta_k}$, with $\pi_j^{(d)}$ the time-series-specific transition distribution and $\theta_k$ the set of HMM emission parameters.  Over the collection of $D$ time series, we center $G^{(1)}_j,\dots,G^{(D)}_j$ around a common \emph{state-$j$-specific transition measure} $G^{(0)}_j$.  Then, each of the infinite collection of state-specific transition measures $G^{(0)}_1,G^{(0)}_2,\dots$ are centered around a global measure $G_0$.  Specifically,
%
\begin{align}                  
	G_0 \sim \DP{\gamma}{H} \qquad G^{(0)}_j\mid G_0 \sim \DP{\eta}{G_0} \qquad G^{(d)}_j\mid G^{(0)}_j \sim \DP{\alpha}{G^{(0)}_j}.
\end{align}
Such a hierarchy allows for more variability between the transition distributions than the specification of Equation~\eqref{eqn:HDPCorporaHierarchy} by only directly coupling state-specific distributions between time series.  The sharing of information between \emph{states} occurs at a higher level in the latent hierarchy (i.e., one less directly coupled to observations).

Although straightforward extensions of existing models, the models presented in this section have not been discussed in the literature to the best of our knowledge.  Instead, typical models for coupling multiple time series, each modeled via an HMM, rely on assuming exact sharing of the same transition matrix.  (In the LDA framework, that would be equivalent to a model in which every document $d$ shared the same topic weights, $\pi^{(d)} = \pi_0$.)  With such a formulation, each time series (\emph{entity}) has the exact same mixed membership with the global collection of dynamic regimes (\emph{attributes}).

Alternatively, models have been proposed in which each time series~$d$ is hard-assigned to one of some $M$ distinct HMMs, where each HMM is comprised of a unique set of states and corresponding transition distributions and emission parameters.  For example, \cite{Qi:07} and~\cite{Lennox:10} examine a Dirichlet process mixture of HMMs, allowing $M$ to be unbounded.  Based on a fixed assignment of time series to some subset of the global collection of HMMs, this model reduces to $M'$ examples of exact sharing of HMM parameters, where $M'$ is the number of unique HMMs assigned.  That is, there are $M'$ clusters of time series with the exact same mixed membership among a set of attributes (i.e., dynamic regimes) that are distinct between the clusters.  

By defining a global collection of dynamic regimes and time-series-specific transition distributions, the formulations proposed above instead allow for commonalities between parameterizations while maintaining time-series-specific variations in the mixed membership.  These ideas more closely mirror the LDA mixed membership story for a corpus of documents.
%
\paragraph{The Beta-Bernoulli Process HMM}
%
%
Analogously to HDP-LDA, the HDP-based models for a collection of (or a single) time series assume that each time series has membership with an infinite collection of dynamic regimes.  This is due to the fact that each transition distribution $\pi_j^{(d)}$ has positive mass on the countably infinite collection of dynamic regimes.  In practice, just as a finite-length document is comprised of a finite set of instantiated topics, a finite-length time series is described by a limited set of dynamic regimes.  This limited set might be related yet distinct from the set of dynamic regimes present in another time series.  For example, in the case of the exercise routines, perhaps one observed individual performs jumping jacks, side twists, and arm circles, whereas another individual performs jumping jacks, arm circles, squats, and toe touches.  In a similar fashion to the feature-based approach of the focused topic model described in Section~\ref{sec:BNP-LDA}, one can employ the beta-Bernoulli process to directly capture a \emph{sparse} set of associations between time series and dynamic regimes.  

The beta process framework provides a more abstract and flexible representation of Bayesian nonparametric mixed membership in a collection of time series.  Globally, the collection of time series are still described by a shared library of infinitely many possible dynamic regimes.  Individually, however, a given time series is modeled as exhibiting some sparse subset of these dynamic regimes.

More formally,~\cite{Fox:NIPS09} propose the following specification: each time series~$d$ is endowed with an infinite-dimensional feature vector $\fset{d} = [f_{d1}, \, f_{d2}, \ldots]$, with $f_{dj}=1$ indicating the inclusion of dynamic regime~$j$ in the membership of time series~$d$.  The feature vectors for the collection of $D$ time series are coupled under a common beta process measure $B \sim \mbox{BP}(c,B_0)$.  In this scenario, one can think of $B$ as defining coin-flipping probabilities for the global collection of dynamic regimes.  Each feature vector $\fset{d}$ is implicitly modeled by a Bernoulli process draw $X_d\mid B \sim \mbox{BeP}(B)$ with $X_d = \sum_k f_{dk} \delta_{\theta_k}$.  That is, the beta-process-determined coins are flipped for each dynamic regime and the set of resulting heads indicate the set of selected features (i.e., via $f_{dk}=1$).  

The beta process specification allows flexibility in the number of total and time-series-specific dynamic regimes, and encourages time series to share similar subsets of the infinite set of possible dynamic regimes.  Intuitively, the shared sparsity in the feature space arises from the fact that the total sum of coin-tossing probabilities is finite and only certain dynamic regimes have large probabilities.  Thus, certain dynamic regimes are more prevalent amongst the time series, though the resulting set of dynamic regimes clearly need not be identical.  For example, the lower subfigure in Figure~\ref{fig:BPBePrealizations}(a) illustrates a collection of feature vectors drawn from this process.

To limit each time series to solely switch between its set of selected dynamic regimes, the feature vectors are used to form \emph{feature-constrained transition distributions}:
\begin{align}
	\symsubsup{\pi}{j}{d} \mid \fset{d} \sim \mbox{Dir}([\gamma,\dots,\gamma,\gamma + \kappa,\gamma,\dots]\otimes \fset{d}).\label{eqn:DirPrior2} 
\end{align}
Again, we use $\mbox{Dir}([\gamma,\dots,\gamma,\gamma + \kappa,\gamma,\dots] \otimes \fset{d})$ to denote a Dirichlet distribution defined over the finite set of dimensions specified by $\fset{d}$ with hyperparameters given by the corresponding subset of $[\gamma,\dots,\gamma,\gamma + \kappa,\gamma,\dots]$.  Here, the $\kappa$ hyperparameter places extra expected mass on the component of $\symsubsup{\pi}{j}{d}$ corresponding to a self-transition $\symsubsup{\pi}{jj}{d}$, analogously to the sticky hyperparameter of the sticky HDP-HMM~\citep{Fox:AOAS11}.  This construction implies that $\symsubsup{\pi}{j}{d}$ has only a finite number of non-zero entries $\symsubsup{\pi}{jk}{d}$.  As an example, if
\begin{align*}
\fset{d}=\begin{bmatrix} 1 & 0 & 0 & 1 & 1 & 0 & 1 & 0 & 0 & 0 \cdots \end{bmatrix},
\end{align*} 
then
\begin{align*}
\symsubsup{\pi}{j}{d} =\begin{bmatrix}\symsubsup{\pi}{j1}{d} & 0 & 0 & \symsubsup{\pi}{j4}{d} & \symsubsup{\pi}{j5}{d} & 0 & \symsubsup{\pi}{j7}{d} & 0 & 0 & 0 \cdots\end{bmatrix}
\end{align*}
with $\begin{bmatrix}\symsubsup{\pi}{j1}{d} & \symsubsup{\pi}{j4}{d} & \symsubsup{\pi}{j5}{d} & \symsubsup{\pi}{j7}{d}\end{bmatrix}$ distributed according to a 4-dimensional Dirichlet distribution.  Pictorially, the generative process of the feature-constrained transition distributions is similar to that illustrated in Figure~\ref{fig:LDAsparsestick}.

Although the methodology described thus far applies equally well to HMMs and other Markov switching processes,~\cite{Fox:NIPS09} focus on the AR-HMM of Equation~\eqref{eqn:SVAR}.  Specifically, let $\symsubsupB{y}{t}{d}$ represent the observed value of the $d$th time series at time $t$, and let $\symsubsup{z}{t}{d}$ denote the latent dynamical regime. Assuming an order-$r$ AR-HMM, we have
\begin{equation}
	\begin{aligned}
		\symsubsup{z}{t}{d} \mid \{\symsubsup{\pi}{j}{d}\},\symsubsup{z}{t-1}{d} &\sim \symsubsup{\pi}{\symsubsup{z}{t-1}{d}}{d}\\
		\symsubsupB{y}{t}{d} &= \sum_{j=1}^r A_{j,\symsubsup{z}{t}{d}}\symsubsupB{y}{t-j}{d} + \symsubsupB{e}{t}{d}(\symsubsup{z}{t}{d}), 
	\end{aligned}
	\label{eqn:multSVAR} 
\end{equation}
where $\symsubsupB{e}{t}{d}(k) \sim \mathcal{N}(0,\Sigma_{k})$. Recall that each of the $\theta_k = \{\BF{A}_{k},\Sigma_k\}$ defines a different VAR($r$) dynamic regime and the feature-constrained transition distributions $\pi^{(d)}$ restrict time series $d$ to transition amongst dynamic regimes (indexed at time $t$ by $\symsubsup{z}{t}{d}$) for which it has membership, as indicated by its feature vector $\fset{d}$.
\begin{figure}
	[t!] \centering \hspace{0.2in} 
	\includegraphics[height=2.5in]{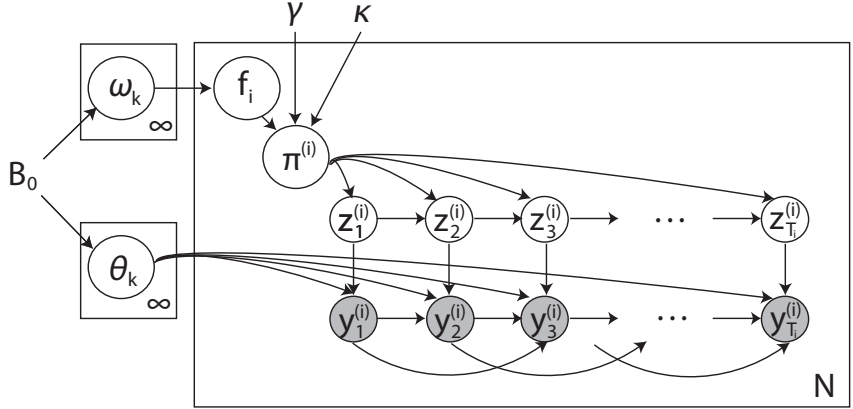}
	\caption[Graphical model of the BP-AR-HMM.] {Graphical model of the BP-AR-HMM. The beta process distributed measure $\mbox{$B \mid B_0 \sim \mbox{BP}(1,B_0)$}$ is represented by its masses $\omega_k$ and locations $\theta_k$, as in Equation~\eqref{eqn:CRM}. The features are then conditionally independent draws $\mbox{$f_{dk} \mid \omega_k \sim \mbox{Bernoulli}(\omega_k)$}$, and are used to define feature-constrained transition distributions $\mbox{$\symsubsup{\pi}{j}{d} \mid \fset{d} \sim \mbox{Dir}([\gamma,\dots,\gamma,\gamma+\kappa,\gamma,\dots]\otimes \fset{d})$}$. The switching VAR dynamics are as in Equation~\eqref{eqn:multSVAR}.} \label{fig:BPARHMM} 
\end{figure}

Conditioned on the set of $D$ feature vectors $\fset{d}$ coupled via the beta-Bernoulli process hierarchy, the model reduces to a collection of $D$ switching VAR processes, each defined on the finite state space formed by the set of selected dynamic regimes for that time series.  Importantly, the beta-process-based featural model couples the dynamic regimes exhibited by different time series.  Since the library of possible dynamic parameters is shared by all time series, posterior inference of each parameter set $\theta_k$ relies on pooling data amongst the time series that have $f_{dk}=1$. It is through this pooling of data that one may achieve more robust parameter estimates than from considering each time series individually.

The resulting model is termed the \emph{BP-AR-HMM}, with a graphical model representation presented in Figure~\ref{fig:BPARHMM}.  The overall model specification is summarized as\footnote{One could consider alternative specifications of $\beta = [\gamma,\dots,\gamma,\gamma + \kappa,\gamma,\dots]$, such as in the focused topic model of Equation~\eqref{eqn:BP-LDA} where each element $\beta_k$ is an independent random variable. Note that~\cite{Fox:NIPS09} treat $\gamma,\kappa$ as random.}:
\begin{equation}
	\begin{aligned}
		B &\sim \mbox{BP}(1,B_0)\\
		X_d \mid B &\sim \mbox{BeP}(B), \quad d = 1,\dots, D\\
		\symsubsup{\pi}{j}{d} \mid \fset{d} &\sim \mbox{Dir}([\gamma,\dots,\gamma,\gamma + \kappa,\gamma,\dots]\otimes \fset{d}), \quad d=1,\dots,D, \,\, j=1,2,\dots\\
		\symsubsup{z}{t}{d}\mid \{\symsubsup{\pi}{j}{d}\},\symsubsup{z}{t-1}{d} &\sim \symsubsup{\pi}{\symsubsup{z}{t-1}{d}}{d}, \quad d=1,\dots,D, \,\, t=1,\dots,T_d\\
		\symsubsupB{y}{t}{d} &= \sum_{j=1}^r A_{j,\symsubsup{z}{t}{d}}\symsubsupB{y}{t-j}{d} + \symsubsupB{e}{t}{d}(\symsubsup{z}{t}{d}),  \quad d=1,\dots,D, \,\, t=1,\dots,T_d.
	\end{aligned}
	\label{eqn:BPARHMM} 
\end{equation}
\cite{Fox:NIPS09} apply the BP-AR-HMM to the analysis of multiple motion capture (MoCap) recordings of people performing various exercise routines, with the goal of jointly segmenting and identifying common dynamic behaviors amongst the recordings.  In particular, the analysis examined six recordings taken from the CMU database~\citep{CMUmocap}, three from Subject 13 and three from Subject 14. Each of these routines used some combination of the following motion categories: running in place, jumping jacks, arm circles, side twists, knee raises, squats, punching, up and down, two variants of toe touches, arch over, and a reach out stretch.  

\begin{figure}[t!] 
	\centering 
	\includegraphics[width = \columnwidth]{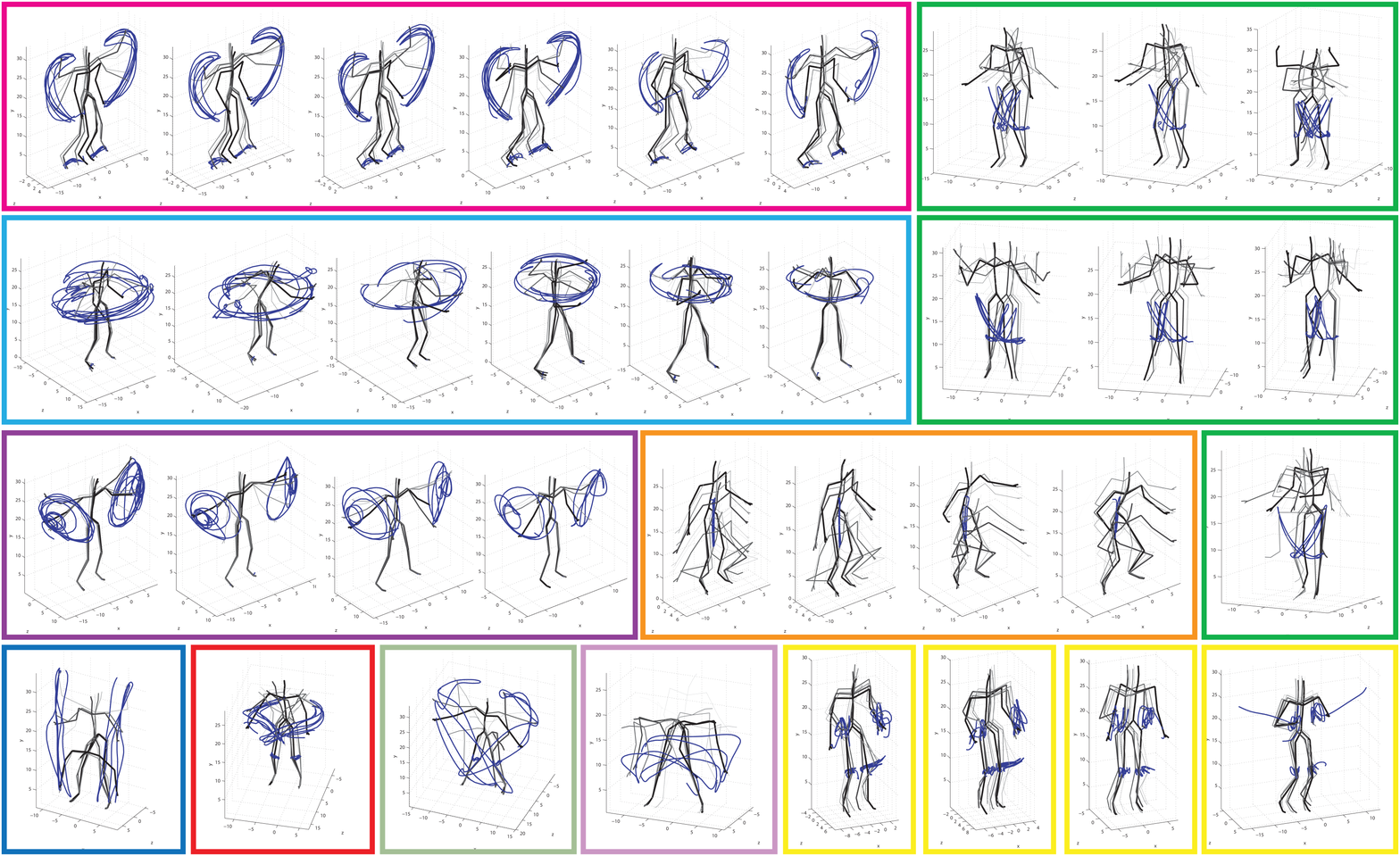} \caption[Motion capture skeleton plots for BP-AR-HMM learned segmentations of six exercise routine videos.] {Each skeleton plot displays the trajectory of a learned contiguous segment of more than two seconds, bridging segments separated by fewer than 300 msec. The boxes group segments categorized under the same behavior label, with the color indicating the true behavior label (allowing for analysis of split behaviors). Skeleton rendering done by modifications to Neil Lawrence's Matlab MoCap toolbox~\citep{LawrenceMoCap}.}\label{fig:MoCap}
\end{figure}

The resulting segmentation from the joint analysis is displayed in Figure~\ref{fig:MoCap}.  Each skeleton plot depicts the trajectory of a learned contiguous segment of more than two seconds, and boxes group segments categorized under the same behavior label in the posterior. The color of the box indicates the true behavior label. From this plot we can infer that although some true behaviors are split into two or more categories (``knee raises'' [green] and ``running in place'' [yellow])\footnote{The split behaviors shown in green and yellow correspond to the true motion categories of knee raises and running, respectively, and the splits can be attributed to the two subjects performing the same motion in a distinct manner.}, the BP-AR-HMM is able to find common motions (e.g., six examples of ``jumping jacks'' [magenta]) while still allowing for various motion behaviors that appeared in only one movie (bottom left four skeleton plots.)

\begin{figure}
	[t!] \centering 
	\includegraphics[width=.7\textwidth]{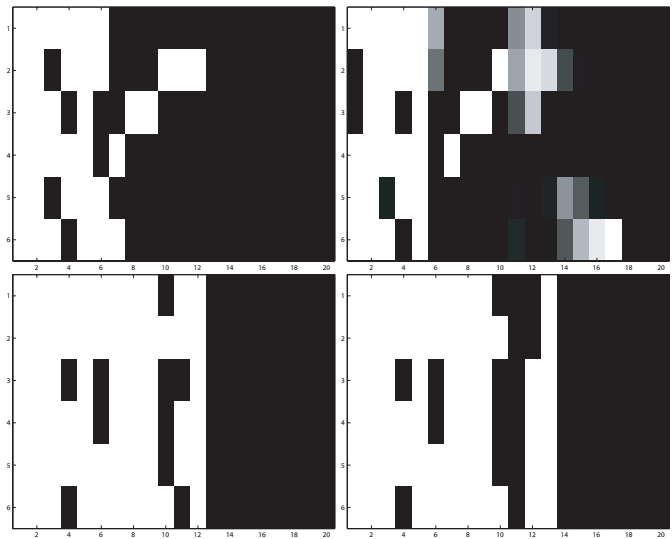} \caption[Learned MoCap feature matrices from the BP-AR-HMM, HMM, and Gaussian mixture model approaches.]{Feature matrices associated with the true MoCap sequences (left) and BP-AR-HMM estimated sequences over iterations 15,000 to 20,000 of an MCMC sampler (right).  Each row is an individual recording and each column a possible dynamic behavior.  The white squares indicate the set of selected dynamic behaviors.} \label{fig:MoCap_F} 
\end{figure}

The key characteristic of the BP-AR-HMM that enables the clear identification of shared versus unique dynamic behaviors is the fact that the model takes a feature-based approach.  The true feature matrix and BP-AR-HMM estimated matrix, averaged over a large collection of MCMC samples, are shown in Fig.~\ref{fig:MoCap_F}.  Recall that each row represents an individual recording's feature vector $\fset{d}$ drawn from a Bernoulli process, and coupled under a common beta process prior.  The columns indicate the possible dynamic behaviors (truncated to a finite number if no assignments were made thereafter.)
%
%
\section{Related Bayesian and Bayesian Nonparametric Time Series Models}
\label{sec:related}
In addition to the regime-switching models described in this article, there is large and growing literature 
on Bayesian parametric and nonparametric time series models, many of which also have interpretations as 
mixed membership models.  We overview some of this literature in this section, aiming not to cover the 
entirety of related literature but simply to highlight three main themes: (i) \emph{non-homogeneous} 
mixed membership models, and relatedly, \emph{time-dependent processes}, (ii) other \emph{HMM-based models}, 
and (iii) time-independent \emph{mixtures of autoregressions}.
\subsection{Non-Homogeneous Mixed Membership Models}
\label{sec:nonhomogeneous}
\subsubsection*{Time-Varying Topic Models}
The documents in a given corpus sometimes represent a collection spanning a wide range of time.  It is likely that the prevalence and popularity of various topics, and words within a topic, change over this time period.  For example, when analyzing scientific articles, the set of scientific questions being addressed naturally evolves.  Likewise, within a given subfield, the terminology similarly develops---perhaps new words are created to describe newly discovered phenomena or other words go out of vogue.

To capture such changes,~\cite{BleiLafferty:06} proposed a \emph{dynamic topic model}.  This model takes the general framework of LDA, but specifies a Gaussian random walk on a set of topic-specific word parameters
\begin{align}
	\theta_{t,k} \mid \theta_{t-1,k} \sim \mathcal{N}(\theta_{t-1,k},\sigma^2 I)
\end{align}
and document-specific topic parameters
\begin{align}
	\beta_{t} \mid \beta_{t-1} \sim \mathcal{N}(\beta_{t-1},\delta^2 I).
\end{align}
The topic-specific word \emph{distribution} arises via $\pi(\theta_{k,t,w}) = \frac{\exp(\theta_{k,t,w})}{\sum_{w} \exp(\theta_{k,t,w})}$.  For the topic distribution,~\cite{BleiLafferty:06} specify $\eta \sim \mathcal{N}(\beta_t,a^2 I)$ and transform to $\pi(\eta)$.  This formulation provides a \emph{non-homogeneous} mixed membership model since the membership weights (i.e., topic weights) vary with time.

The formulation of~\cite{BleiLafferty:06} assumes a discrete, evenly spaced corpora of documents.  Often, however, documents are observed at non-evenly and potentially finely sampled time points.  \cite{WangBlei:08} explore a continuous time extension by modeling the evolution of $\theta_{t,k}$ as Brownian motion.  As a simplifying assumption, the authors do not consider evolution of the global topic proportions $\beta$.

\subsubsection*{Time-Dependent Bayesian Nonparametric Processes}
For Bayesian nonparametric time-varying topic modeling,~\cite{Srebro:05} propose a time-dependent Dirichlet process.  The Dirichlet process allows for an infinite set of possible topics, in a similar vein to the motivation in HDP-LDA.  Importantly, however, this model does \emph{not} assume a mixed membership formulation and instead takes each document to be hard-assigned to a single topic.  The proposed time-dependent Dirichlet process models the changing popularity of various topics, but assumes that the topic-specific word distributions are static.  That is, the Dirichlet process probability measures have time-varying weights, but static atoms.

More generally, there is a growing interest in time-dependent Bayesian nonparmetric processes.  The dependent Dirichlet process was originally proposed by~\cite{MacEachern:99}.  A substantial focus has been on evolving the weights of the random discrete probability measures.  Recently,~\cite{GriffinSteel:11} examine a general class of autoregressive stick-breaking process, and~\cite{Mena:11} study stick-breaking processes for continuous-time modeling.  \cite{Taddy:10} considers an alternative autoregressive specification for Dirichlet process stick-breaking weights, with application to modeling the changing rate function in a dynamic spatial Poisson process.
%
\subsection{Hidden-Markov-Based Bayesian Nonparametric Models}
A number of other Bayesian nonparametric models have been proposed in the literature that take as 
their point of departure a latent Markov switching mechanism.
Both the infinite factorial HMM~\citep{VanGael:08} and the infinite hierarchical HMM~\citep{Heller:09} provide Bayesian nonparametric priors for infinite collections of latent Markov chains.  The infinite factorial HMM provides a distribution on \emph{binary} Markov chains via a \emph{Markov Indian buffet process}.  The implicitly defined time-varying infinite-dimensional binary feature vectors are employed in performing blind source separation (e.g., separating an audio recordings into a time-varying set of overlapping speakers.)  The infinite hierarchical HMM also employs an infinite collection of Markov chains, but the evolution of each depends upon the chain above.  Instead of modeling binary Markov chains, the infinite hierarchical HMM examines finite multi-class state spaces.

Another method that is based on a finite state space is that of~\cite{Taddy:09}.  The proposed model assumes that each HMM state defines an independent Dirichlet process regression.  Extensions to non-homogenous Markov processes are considered based on external covariates that inform the latent state.

In~\cite{SaeediBouchard:11}, the authors propose a hierarchical gamma-exponential process for modeling recurrent continuous time processes.  This framework provides a continuous-time analog to the discrete-time sticky HDP-HMM.

Instead of Markov-based regime-switching models that capture repeated returns to some (possibly infinite) set of dynamic regimes, one can consider changepoint methods in which each transition is to a new dynamic regime.  Such methods often allow for very efficient computations.  For example,~\cite{Xuan:07} base such a model on the product partition model\footnote{A product partition model is a model in which the data are assumed independent across some set of unknown partitions~\citep{Hartigan:90,BarryHartigan:92}.  The Dirichlet process is a special case of a product partition model.} framework to explore changepoints in the dependency structure of multivariate time series, harnessing the efficient dynamic programming techniques of~\cite{Fearnhead:06}.  More recently,~\cite{Zantedeschi:11} explore a class of dynamic product partition models and online computations for predicting movements in the term structure of interest rates.
\subsection{Bayesian Mixtures of Autoregressions}
In this article, we explored two forms of switching autoregressive models: the HDP-AR-HMM and the BP-AR-HMM.  Both models assume that the switches between autoregressive parameters follow a discrete-time Markov process.  There is also substantial literature on nonlinear autoregressive modeling via mixtures of autoregressive processes, where the mixture components are independently selected over time.  \cite{LauSo:08} consider a Dirichlet process mixture of autoregressions.  That is, at each time step the observation is modeled as having been generated from one of an unbounded collection of autoregressive processes, with the mixing distribution given by a Dirichlet process.  A variational approach to Dirichlet process mixtures of autoregressions with unknown orders has recently been explored in~\cite{Morton:11}.  \cite{WoodRosenKohn:11} aim to capture the idea of \emph{structural breaks} by segmenting a time series into contiguous blocks of $L$ observations and assigning each \emph{segment} to one of a finite mixture of autoregressive processes; implicitly, all $L$ observations are associated with a given mixture component.  Key to the formulation is the inclusion of time-varying mixture weights, leading to a nonstationary process as in Section~\ref{sec:nonhomogeneous}.  

As an alternative formulation that captures Markovianity, but not directly in the latent mixture component,~\cite{MullerWestMacEachern:97} consider a model in which the probability of choosing a given autoregressive component is modeled via a kernel based on the previous set of observations (and potential covariates).  The maximal set of $K$ mixture components is fixed, with the associated autoregressive parameters taken to be draws from a Dirichlet process, implying that only $k\leq K$ will take distinct values.
\section{Discussion}
In this article, we have discussed a variety of time series models that have interpretations in the mixed membership framework.  Mixed membership models are comprised of three key components: \emph{entities}, \emph{attributes}, and \emph{data}.  What differs between mixed membership models is the type of the data associated with each entity, and how the entities are assigned membership with the set of possible attributes.  Abstractly, in our case each time series is an \emph{entity} that has membership with a collection of dynamic regimes, or \emph{attributes}.  The partial memberships are determined based on the temporally structured observations, or \emph{data}, for the given time series.  This structured data is in contrast to the typical focus of mixed membership models on exchangeable collections of data per entity (e.g., a bag of words representation of a document's text.)

Throughout the article, we have focused our attention on the class of \emph{Markov switching processes}, and further restricted our exposition to Bayesian parametric and nonparametric treatments of such models.  The latter allows for an unbounded set of attributes by modeling processes with Markov transitions between an infinite set of dynamic regimes.  For the class of Markov switching processes, the mixed membership of a given time series is captured by the time-series-specific set of Markov transition distributions.  Examples include the classical hidden Markov model (HMM), autoregressive HMM, and switching state-space model.  

In mixed membership modeling, one typically has a group of entities (e.g., a corpus of documents) and the goal is to allow each entity to have a unique set of partial memberships amongst a shared collection of attributes (e.g., topics).  Through such modeling techniques, one can efficiently and flexibly share information between the data sources associated with the entities.  Motivated by such goals, in this article we explored a nontraditional treatment of time series analysis by examining models for \emph{collections} of time series.  We proposed a Bayesian nonparametric model for multiple time series based on ideas analogous to Dirichlet-multinomial modeling of documents.  We also reviewed a Bayesian nonparametric model based on a beta-Bernoulli framework that directly allows for sparse association of time series with dynamic regimes.  Such a model enables decoupling the presence of a dynamic regime from its prevalence.  

The discussion herein of time series analysis from a mixed membership perspective has been previously neglected, and leads to interesting ideas for further development of time series models.
	
%

\bibliographystyle{chicago}
\bibliography{../../Bibliography/Bibliography_BPARHMM}

\end{document}